\documentclass[%
 aip,
 apl,%
 amsmath,
 amssymb,
preprint,%
]{revtex4-1}

\usepackage[hidelinks]{hyperref}
\usepackage{graphicx}
\usepackage{dcolumn}
\usepackage{bm}

\usepackage[version=4]{mhchem}
\usepackage{chemformula}
\usepackage{siunitx}
\DeclareSIUnit{\torr}{Torr}
\usepackage[caption=false]{subfig}
\usepackage{floatrow}
\usepackage{cleveref}
\begin{document}

\title{A fabrication guide for planar silicon quantum dot heterostructures}

\author{Paul C. Spruijtenburg}%
\author{Sergey V. Amitonov}
\author{Wilfred G. van der Wiel}
\author{Floris A. Zwanenburg}

\affiliation{ 
NanoElectronics Group, MESA$^+$ Institute for Nanotechnology, University of Twente, P.O. Box 217,
7500 AE Enschede, The Netherlands
}%

\date{\today}

\begin{abstract}
  We describe important considerations to create top-down fabricated planar quantum dots in silicon, often not discussed
  in detail in literature.  The subtle interplay between intrinsic material properties, interfaces and fabrication
  processes plays a crucial role in the formation of electrostatically defined quantum dots. Processes such as
  oxidation, physical vapor deposition and atomic-layer deposition must be tailored in order to prevent unwanted side
  effects such as defects, disorder and dewetting.  In two directly related manuscripts written in parallel we use
  techniques described in this work to create depletion-mode quantum dots in intrinsic silicon, and low-disorder silicon
  quantum dots defined with palladium gates.  While we discuss three different planar gate structures, the general
  principles also apply to 0D and 1D systems, such as self-assembled islands and nanowires.
\end{abstract}

\maketitle

Dealing with the fragility of the quantum coherent state is one of the key issues on the road to meet the limits posited
by quantum computation schemes~\cite{Fowler2012}. It is the coupling of quantum states to states in an unknown
environment which is the driver for decoherence. The properties of the environment therefore dictate the performance of
a quantum bit (qubit).  Creating qubits in the solid state means that the environment consists of many different
materials and structures used in device construction. Quantum dots have been created by confining carriers in
e.g. AlGaAs/GaAs~\cite{Chen2012} or Si/SiGe heterostructures~\cite{Borselli2011a}, to dopant atoms in a host material,
in SiGe nanohuts~\cite{Katsaros:2010jz}, Si and Si/Ge nanowires~\cite{Zwanenburg2009,brauns2016highly}, etched mesas on
silicon-on-insulator~\cite{Horibe2015}, and planar MOS structures~\cite{Lim2009}.  The intrinsic properties of these
(material) systems give rise to interactions detrimental to qubit creation and readout.

The hyperfine interaction of nuclear spins in the host material and the qubit is one such effect. The non-zero-spin isotopes in a
material create a nuclear-spin-bath and cause decoherence of the quantum state. This is the motivation for
the use of isotopically purified silicon as a host material for spin qubits~\cite{Zwanenburg2013,DeSousa2003}. The
purified silicon, now containing predominantly \ce{^{28}Si}, results in a zero-nuclear-spin isotope system, and
eliminates the fluctuations in the spin bath, which are detrimental.

Field noise, such as charge- and spin-noise~\cite{Kuhlmann2013} can also influence the lifetime of the quantum
state. One strategy to deal with these fluctuations is to tune the quantum dots to certain regimes in phase space where
the energy levels of interest are insensitive to these fields. This can happen in the clock transition of Bi dopants in
Si~\cite{Rashidi2016a}, by dressing qubits and tuning them appropriately~\cite{Laucht2016}, or for hybrid quantum dots. ~\cite{Shi}
Another effect that can influence quantum states are fluctuations in the electrochemical potential at longer timescales. These
have been shown to occur due to charge offsets fluctuating over time in glassy media and their intrinsic two-level
systems (TLS).~\cite{Zimmerman2008,Stewart2016}

Finally, unintentional quantum dots, charge traps, or charge defects can influence a desired quantum state. This class
of effects can manifest when the scale of the wave function of the charge carriers involved is equal to the scale of
(unintended) features or variations in the structure. In silicon, the electrons have a larger transverse effective mass
($m_t^* = 0.19m_0$) as compared to e.g. the electrons in \ce{GaAs} ($m^* = 0.067m_0$). The electrons in silicon therefore have a smaller
wave packet. Smaller features in the structure or atomic scale defects can thus play a more significant role.

The definition and the quality of the quantum state is not only determined by the host material in which the quantum
state mostly resides. The entire heterostructure interacts with the quantum state through various effects.
Therefore, great care must be taken in the creation of the heterostructure and materials, in order to realize an ideal
quantum dot. 

We focus here exclusively on considerations as they pertain to quantum computation in a planar silicon quantum dot
structure, but the mechanisms found do not exclusively pertain to this particular device type. Many of the identified
issues come about during fabrication or find their origin in fundamental material properties and are thus applicable
over a wide range of structures made in the solid state.  Also, this article is strictly limited to planar quantum dots
fabricated for electron transport experiments and does not cover quantum dots made for optical spectroscopy
measurements. Optical spectroscopy on quantum dots and ensembles of dopants is a very active field; see, for example,
the recent work by \textcite{greenland2010coherent}, \textcite{steger2012quantum}, \textcite{Dohnalova}, and references therein. Other quantum
dot systems, such as ensembles of quantum dots, dopants, or colloidal quantum dots are beyond the scope of this article.

This article has the intention to provide a foothold for entrants in the field (e.g. starting graduate students) and
elucidate mechanisms that need to be taken into account when designing and fabricating these devices in the solid
state. It can be read back to front, and may also serve well to be read in an encyclopedic manner.

Criteria for carrier confinement of good quality in planar quantum dots are e.g. charge stability, long spin lifetimes,
and the absence of unintentional quantum dots. Based on these aspects we will look at the heterostructure and its
fabrication, and work out which effects are of importance.

To this end, we will first introduce the planar heterostructure, as a means of effecting quantum dots electrostatically
in a semiconductor. We will then introduce the three heterostructures used in this work, based on the planar
heterostructure.  Then we will briefly discuss relevant layer growth techniques and heterostructure creation in
general. Next, we will review the Si planar quantum dot heterostructure in its entirety, and identify key points in the
heterostructure where effects might occur which are deleterious to quantum dot quality. Finally, we will take an
in-depth look at the role of annealing and hydrogen in these heterostructures.  In two directly related manuscripts
written and submitted at the same time we use recipes from this cookbook to create depletion-mode quantum dots in
intrinsic silicon~\cite{amitonov2017}, and low-disorder silicon quantum dots defined with palladium
gates~\cite{brauns2017}.

\section [The electrostatically defined quantum dot heterostructure]{The electrostatically defined quantum~dot~heterostructure}\label{sec:basic_heterostructure}
The general working principle of the planar quantum dot heterostructure closely resembles that of a MOSFET -- in that
the band structure of the semiconductor is manipulated electrostatically by local gate electrodes, which are separated
electrically from the silicon underneath by an isolating layer. The electrodes control the flow of carriers in the
device and shape the potential profile required for quantum dot formation. The lateral configuration of
the electrodes determines the region where the potential can be manipulated, while the voltage applied to individual gate
electrodes creates the electric field that can control the height of the potential.  When the voltage applied to the
gate electrodes is sufficient, the conduction band (CB) or valence band (VB) is pulled above or below the Fermi level so
that states become available for transport, confining the carriers to a 2D carrier gas.

In order to measure conductance through the device, source and drain regions are created by doping and ohmic contacting.
Implantation regions form source and drain reservoirs and can be ohmically contacted to measure conductance through the device. 
Figure~\ref{fig:regular_hetero} shows the ideal case of the resulting potential in the heterostructure profile, being two
tunnel barriers and a quantum dot in the middle, for the applied voltages on the gate electrodes.

\subsection{Fabrication strategy}
\begin{figure}[htp]
\subfloat{
  \includegraphics[width=.4\linewidth]{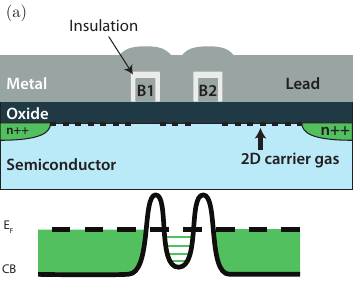}
  \label{fig:ebl_side}
}
\subfloat{
    \includegraphics[width=.4\linewidth] {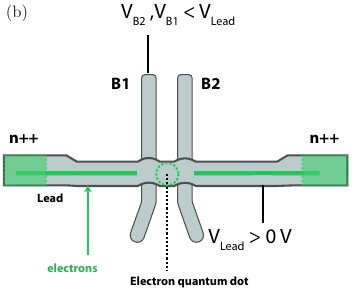}
  \label{fig:ebl_top}
}
\centering

\subfloat{
  \includegraphics[width=.6\linewidth]{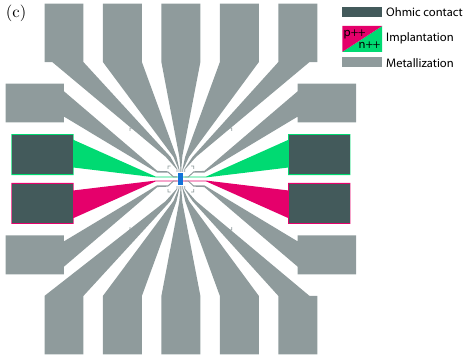}
  \label{fig:photolitho}
}
\caption{ Planar quantum dot heterostructure. \protect\subref{fig:ebl_side} a cross-sectional view for an n-type
  implantation with the desired potential profile of a quantum dot for electrons. \protect\subref{fig:ebl_top} a
  top-view.  \protect\subref{fig:photolitho} Chip design at mesoscale with a design area of
  $2 \times 2$~$\SI{}{\mm\squared}$, showing the implantation regions in red and green for holes and electrons,
  respectively, ohmic contacts in dark grey and the gate architecture in light grey. The center of the device area (dark
  blue) is comprised of high-quality oxide (approximately $70 \times 30$~$\SI{}{\micro\meter\squared}$).  }
  \label{fig:regular_hetero}
\end{figure}

All devices in this work are made using the same fabrication strategy: first, the ``big'' structures, down to
approximately 2 micron resolution, are defined at wafer scale using photolithography. This entails the oxidation of Si
to form \ce{SiO2}, the formation of the implantation regions for the source and drain contacts, creation of the
(high-quality) oxide, and the metallization of the electrodes which can be contacted later for
measurement. Figure~\ref{fig:photolitho} shows the top view of the structures created in this process. Then, the smaller
structures can be defined. Electron-beam lithography (EBL), followed by evaporation and lift-off is used to define the
sub-micron gate electrodes. The device usually consists of more than one layer of electrodes, and therefore an
insulating layer is needed for electrical separation of electrode layers.  Dividing the fabrication process in two steps
at different scale allows for rapid iteration of device configurations, which are otherwise limited by the speed at
which the definition of gate electrodes with EBL takes place.

\subsection{The ambipolar concept}\label{sec:ambipolar_concept}
The behavior of charge carriers in the valence and conduction-band is different, and both charge carriers have different
properties. As an example, in bulk and at band-minimum, the longitudinal and transverse effective masses for transport of 
electrons are reported as $m^*_{e,l} =0 .98 m_0$, $m^*_{e,t} =0 .19 m_0$, while for holes the effective masses are
$m^*_{hh} = 0.49 m_0$, $m_{lh} = 0.16 m_0$ \cite{van2004principles}, respectively, with $m_0$ the rest mass of the electron.
Furthermore, the spin-orbit coupling for holes in Si is higher than for their counterparts in the conduction band, since
the orbital angular momentum is non-zero. The criteria for quantum dot formation therefore also differ.

\begin{figure}[htp]
\subfloat[]{
    \includegraphics[width=0.6\linewidth]{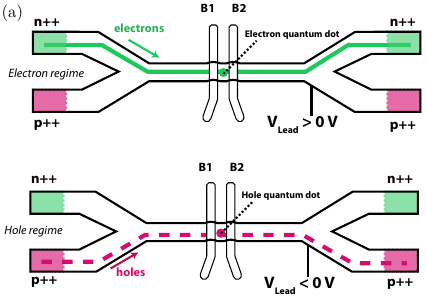}
    \label{fig:ambipolar_top}
}
\subfloat[]{
    \includegraphics[width=0.4\linewidth]{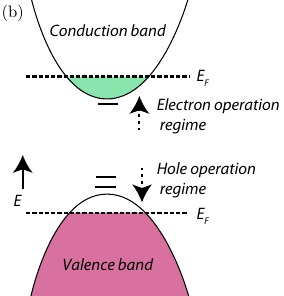}
    \label{fig:ambipolar_bandgap}
}
  \caption{The ambipolar device concept. \protect\subref{fig:ambipolar_top} the two modes of operation of the ambipolar
    device. By applying a negative voltage holes can flow from the source and drain implantation region. A
    positive voltage allows electrons to be used as charge carriers. \protect\subref{fig:ambipolar_bandgap} Schematic
    overview of the bandgap of silicon, including possible states in the bandgap. Not shown here is the indirect nature
    of the bandgap. The range of operation of the ambipolar device is apparent, allowing probing of states close to the
    valence band or conduction band, depending on the operation mode of the device.}
  \label{fig:ambipolar}
\end{figure}

To explore this, ambipolar devices can be created where both holes and electrons can be used as charge
carriers. \cite{Mueller2015a,Betz2014, Jarillo-Herrero2004,Guttinger2009,Pfleiderer1986,Chen2012} The planar quantum dot
design is easily modified to create  p- and n-type implantation regions at either end of the device. Which carrier
is used to transport charge is controlled by gating the intrinsic silicon (Figure~\ref{fig:ambipolar_top}). A negative potential on the gate will pull
the VB above the Fermi level, making hole states available for transport. Conversely, applying a positive potential on
the gate will push the CB below the Fermi level and make electron states available for transport. We call these modes of
operation the hole and electron operation regime (Figure~\ref{fig:ambipolar_bandgap}).  The ability to transport either electrons or holes in the same
crystalline environment allows for the probing of states close to either the CB or VB of the semiconductor, while
discounting fabricational variations that would otherwise exist when measuring these effects from one device to the next.

\subsection{Heterostructure variations}\label{sec:3hetero}
In this work, we will encounter three variations on the basic heterostructure discussed in
\cref{sec:basic_heterostructure}. The different designs attempt to address several issues, which we will discuss later.

\begin{figure}[htp]
  \centering
  \includegraphics[width=\linewidth ]{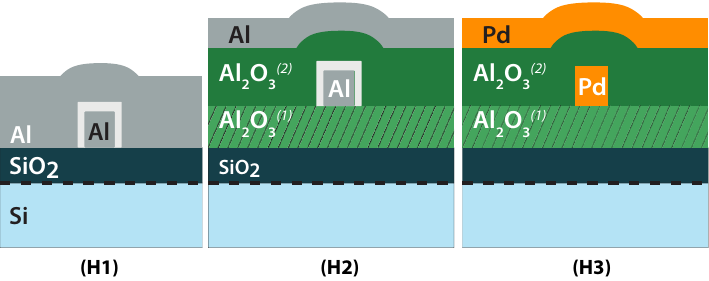}
  \caption{Three heterostructures with (H1) the Al-gate standard architecture (H2) the architecture including
    additional, separately grown, \ce{Al2O3} layers  \textsl{(1),(2)} and (H3) the architecture using palladium as a gate
    electrode material. The source and drain implanted regions have been omitted here.}
  \label{fig:all_hetero}
\end{figure}

All heterostructures use \ce{SiO2} as the first insulating layer.  The first heterostructure (H1) uses Al as the
material for gate electrodes, based on the recipe introduced by Angus \textit{et al.}\cite{Angus2007} and can be seen in
Figure~\ref{fig:all_hetero}(H1). Aluminum has the advantage that an insulating layer is easily created between the
electrodes by thermal oxidation.  In the second heterostructure (H2), a \SI{5}{\nm} layer of \ce{Al2O3} grown at
\SI{250}{\celsius} is added between the \ce{Al} and the \ce{SiO2}, and a capping layer of \ce{Al2O3} (not shown) is
finally grown at \SI{100}{\celsius}.  The third heterostructure (H3) substitutes the \ce{Al} for \ce{Pd} as the
gate-electrode material, and maintains the inter- and capping-layer of \ce{Al2O3}.

\section{Layer synthesis techniques}\label{sec:layer_synthesis}
For a well-defined electrochemical potential profile, it is essential to control the properties of the materials
constituting a heterostructure. Here we will briefly touch on some aspects of these techniques, which
we will relate in \cref{sec:hetero} to how they might influence QD formation.

    \subsection{Physical vapor deposition}
    There are several methods of physical vapor deposition in common use today.

    Firstly, thermal evaporation utilizes resistive heating of a crucible (typically made from W or Mo) filled with the material
    to be evaporated.  For evaporation of metals, there are a few parameters of note. First, the rate of evaporation,
    controlled primarily by temperature, determines the grain size of the thin film. The difference in temperature
    between the evaporated material and the substrate determines the grain size of the polycrystalline film. The second
    aspect is the purity of the evaporated material. Since the evaporated material is situated in a thermally heated
    crucible, there is the possibility of contamination from the crucible material.  For deposition pressures of
    $\num{1e-6} < p_\mathrm{dep} < \SI{1e-2}{\torr}$ the evaporated material can be considered to be in the ballistic
    regime.\cite{lakhtakia2005sculptured}

    The second common technique, electron-beam evaporation, has several advantages in contrast to thermal evaporation,
    in that there is less danger of contamination by the crucible. Because for most materials the e-beam heats the
    sample locally, a small puddle is formed inside the target material. This puddle is thus only in contact with the
    same constituent material.  Putting e-beam evaporation at a disadvantage however, is that X-rays are generated in
    the impact of the high-energy electrons with the target material. Possible damage to the substrate is therefore a
    concern.  Grain size in this system is also controlled by the substrate temperature. Exploiting this, it has been shown for Al
    that cooling of the substrate can greatly improve the grain size of the thin film. \cite{Muller2015} For typical
    evaporation rates of around 1-5 \AA/s grain sizes of 15-40~nm have been reported for Pd~\cite{Amin-Ahmadi2013},
    while for Al the grain sizes vary from 20-40~nm for the same range of rates\cite{Bordo2012}.
        
    Thirdly, atomic layer deposition (ALD) is a technique independently discovered in the 60s in the Soviet Union and in
    the 70s in Finland\cite{Puurunen2014}. It makes possible the controlled layer-by-layer growth of oxides,
    metals, fluorides, nitrides, sulfides, carbides, etc. ALD is a self-limiting process and is a subset of
    chemical vapor deposition.  Its main advantages are the low temperatures at which the process can be used,
    precise control over film thickness, and conformal coverage of the resulting film. In this work ALD is used for the growth of \ce{Al2O3}. 

    In general two precursor gases are used, which react in two separate steps as described in reaction
    equations \ref{eq:ald_sorb}, \ref{eq:ald_oxidize}. 
    
    \begin{align}
      \label{eq:ald_sorb}
      \ch{Al(CH3)3 (g) & + Si-O-H (s) & -> & Si-O-Al(CH3)2 (s)  & +  CH4  (g)\\
      2 H2O (g) & +  Si-O-Al(CH3)2 (s) & -> &  Si-O-Al(OH)2 (s) & + 2 CH4  (g) } \label{eq:ald_oxidize}
    \end{align}
    
    The simplified reaction equations as given in equation~\ref{eq:ald_oxidize} bely some more complicated
    reactions that can take place, as the use of \ce{H2O} as a precursor gas has been shown to create hydrogen
    as a reaction by-product in the ALD-cycle. \cite{Dingemans2010} This causes ALD-grown \ce{Al2O3} to have
    hydrogen content at 2-3 at.\%.  Experiments with deuterated water as a precursor show deuterium diffusing
    from the \ce{Al2O3} film toward the Si interface at annealing temperatures of \SI{400}{\celsius}, where
    dangling bonds are passivated and lead to a lower density of interface states $D_{it}$.

    The growth temperature of the \ce{Al2O3} determines the density of the oxide. The hydrogen content will be
    shown to have an important consequence later on in \cref{sec:annealing}.

    \subsection{\ce{SiO2} growth}\label{sec:si_oxidation}
    The growth of \ce{SiO2} can be accomplished by simply introducing an oxidant to silicon at high temperature. The oxidant
    diffuses through the silicon and oxidizes the silicon from the top up to the desired depth. 

    Wet oxidation is generally used to create thick, but qualitatively poor, oxides at a rate of 400~nm/h at
    \SI{1000}{\celsius} and takes place by introducing \ce{H2O} as the oxidant. Dry oxidation occurs in
    \ce{O2}-containing ambient and oxidizes the silicon at rates far slower than for wet oxidation ($<$ 100~nm/hr). The
    latter process yields a denser oxide with fewer defects and higher breakthrough voltages.

    Oxygen, introduced at high temperature, will oxidize the top layer of the silicon first. The newly created oxide
    takes up more volume than was occupied by silicon before oxidation. This decrease in density allows more oxygen to
    penetrate deeper into the silicon, and oxidizes at the newly formed Si/\ce{SiO2} interface. The process continues
    and the Si/\ce{SiO2} oxidation front progresses further down into the substrate. Because of the increase in volume,
    taking place at \SI{900}{\celsius}, the silicon lattice will be uniformly stressed by the
    \ce{SiO2}. \cite{Thorbeck2014}

    The Deal-Grove model describes the oxidation reaction as occurring at the Si/\ce{SiO2} oxidation front. It describes
    the diffusion of the oxidant through the already formed oxide in terms of concentration differences, for which the
    model assumes steady-state conditions. Discrepancies of the Deal-Grove model for thin oxides have led to new
    insights and it has been shown that a more complete model involving Si emission from the surface describes thin
    oxides better. \cite{Kageshima,Uematsu2002a} The quality of the oxide can be further improved by the addition of
    chlorine-containing gas during oxidation. The reduction of atomic silicon emission from the silicon by the chlorine
    is most likely responsible for this. \cite{Uematsu2002}
    
    It is important to note that the oxidation process is not isotropic. Should thermal oxidation start at a surface
    that is made atomically flat; that flatness would not propagate to the Si/\ce{SiO2} interface under normal
    conditions. Furthermore the conditions for oxidation lead to a wide array of interface morphologies
    \cite{Yamabe2009,Yamabe2001, Ravindra1987}. One possible solution for this might be a new technique using a
    microwave-excited high-density plasma with low electron temperature\cite{Ohmi2006,Aratani2008}. The free radicals
    created in this process have been shown to lead to a more abrupt \ce{Si}/\ce{SiO2} interface and to preserve the
    atomically flat nature of as-prepared wafers.\cite{Kuroda2009} The nature of this interface has not been fully
    explored yet, save for theoretical studies that suggest a phase of atomically flat, and stable
    \ce{SiO2} can exist.\cite{Tsetseris2006}

\section{Defects in quantum dot heterostructures}\label{sec:hetero}
As mentioned previously, the fragile nature of quantum states means that many effects which do not affect mesoscopic
devices, play a big role in nanodevices. 

We identify roughly three categories for heterostructures. These categories will serve as rough guides to determine
where effects occur in heterostructures, which in turn might influence design considerations.  The first category
consists of the inherent material properties of the constituents of the heterostructure. These are mainly determined by
the initial growth process, but can also be influenced during processing.  The second category concerns interfaces,
which can manifest a-priori unexpected phenomena. This is a broad and active area of research unto itself which has
garnered interest due to its fundamental physical concepts involving e.g. the breaking of symmetries.

The third are morphological effects. We can think e.g. of the morphology of the gate layout intended to form intentional
quantum dots. However, there are many more unforeseen morphological effects which can occur and must be considered.

We will look at these categories, guided by the three main heterostructures used in this work, as introduced in \cref{sec:3hetero}.

\subsection{Materials}
We will start by examining the basic building blocks of heterostructures, their materials. It is the intrinsic material
properties of silicon that make it a highly valued candidate for quantum computing. Let us briefly review what material
properties influence the formation of our quantum dots.

        \subsubsection{High-quality oxide}
        The creation of an inversion layer in silicon requires the application of a voltage to electrodes typically
        separated from the silicon by only a few tens of nanometers of oxide. The electric fields over this short distance are
        thus large, and on the order of MV/cm. The oxide should be able to withstand these high voltages and avoid
        pinhole defects that occur when reaching the so-called breakthrough voltage.
        
        For all three heterostructures in this work, the process for creating a high-quality oxide is the same. The
        addition of chlorine during growth improves our breakthrough voltage of the oxide. Without it, significantly smaller
        breakthrough voltages were observed after dopant activation in a rapid thermal annealing process (RTA).

        \subsubsection{Damage incurred by processing}
        However perfect the material grown during a process step, it is possible that high-energy processes during
        further processing of the heterostructure cause damage. The \textit{thermal budget} is a common indicator used
        to signify the maximum allowed thermal load before a device will be too damaged. 
        Process damage manifesting as an increase in defect density is also known to occur when using electron-beam
        evaporation\cite{Schiele}. This is most likely caused by the X-rays generated during the evaporation process. 
        Another defect known to be generated by high-energy radiation is the E$'$ defect in \ce{SiO2}\cite{Messina2009}.

        Thus far, little research has been done on the damage electron beam lithography can cause in
        heterostructures. It is however reasonable to expect that the high-energy electrons (as high as 100 keV) do
        interact with the materials in the heterostructure.  It was shown recently that shallow defects can result
        from exposure to EBL. The resultant degraded mobility could be recovered to a great degree by annealing in
        forming gas (5\% \ce{H2}, \SI{435}{\celsius}) for 25 min.\cite{Kim2017}

        It is not a-priori clear if a high-energy beam or a low-energy beam would be preferable, since scattering
        mechanisms in the entire heterostructure determine where the majority of energy of the beam is absorbed. Several
        models for this type of problem, based on energy deposition in (amorphous) solids have been
        suggested.\cite{Kobetich1968, Kanayat1972} Low-energy patterning (up to a few keV) on thin layers of resist
        could alleviate this problem.

        All heterostructures in this work have been exposed to EBL processes at energies of typically 20 to 28 keV.

        \subsubsection{Dewetting}\label{sec:dewetting}

\begin{figure}[htp]
                   \centering
                   \includegraphics[width=\linewidth]{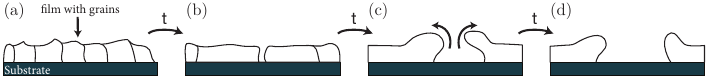}
                   \caption{Schematic cross-sectional overview of the dewetting progress. (a) The start of the
                   dewetting process, with the grains and grain boundaries in the film depicted as-deposited. (b) In the next time-step, the
                   grains have agglomerated and formed bigger domains, with the first gaps forming due to
                   surface tension. (c) A mass flux occurs as mass is transported as indicated by the arrows, and the
                   edge of the material retracts, widening the gap between the wetted regions. (d)
                   The process continues until an equilibrium is reached.}
                          \label{fig:dewetting}
        \end{figure}

        Dewetting is the breaking of continuity of an initially continuous film caused by differences in
        free surface energy at liquid-liquid or solid-liquid interfaces. In our case the interface
        consists of either Al/\ce{SiO2}, Al/\ce{Al2O3} or Pd/\ce{Al2O3}. At elevated temperatures thin films of metal, while
        still below their melting point, can become mobile.  Dewetting becomes energetically favourable under
        the right circumstances: temperature, granularity of the film, free surface energy at the
        Al/$\textrm{Si}\textrm{O}_2$ and vacuum/Al interface. The process is illustrated in Figure~\ref{fig:dewetting}.
        
                \begin{figure}[ht]
                  \centering
                        \subfloat{
                        \includegraphics[width=.5\linewidth]{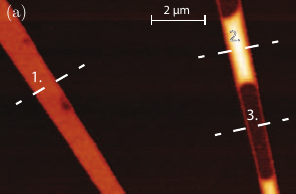}
                        \label{fig:dewetting_afm}
                        }
                        \subfloat{
                        \includegraphics[width=.5\linewidth]{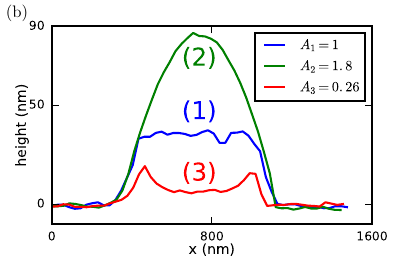}
                        \label{fig:dewetting_profile}
                        }
                \caption{Dewetting behavior of pure aluminum on \ce{SiO2}. \protect\subref{fig:dewetting_afm} AFM image
                  overview of a device after annealing at \SI{400}{\celsius}. \protect\subref{fig:dewetting_profile}
                  Height at the various linecuts in \protect\subref{fig:dewetting_afm} and the areas underneath each
                  curve normalized to $A_1$, the area under curve (1).}

                \end{figure}
                
                Figure~\ref{fig:dewetting_afm} shows dewetting behavior as observed in one of our samples after
                annealing. The left Al electrode, labeled (1) still has its as-deposited height of 36~nm. The right electrode
                shows clear signs of dewetting. At (2) the height has increased from 36~nm as-deposited to a maximum of
                roughly 90~nm. The material for this height increase has been moved from (3) to (2).  The areas $A_i$
                underneath curves taken at (1), (2), and (3) are normalized to $A_1$. We see that
                $ A_2 + A_3 \simeq 2A_1$, i.e the material ``missing'' from (3) has moved to (2), and thus the sum
                equals twice the original.  At the interface region between (2) and (3) the convex/concave shape is
                reminiscent of the classical meniscus in e.g. water, indicating energy minimization caused by surface
                tension. \cite{Thompson2012} This can also be observed on both sides of the electrode at (3).

        \subsubsection{Properties of Al at elevated temperatures} 
        At elevated temperature other effects concerning Al can arise. It has been reported that \ce{AlO_x} hillocks are
        created when exposing Al on \ce{Al2O3} to a high temperature step of \SI{600}{\celsius}
        \cite{Dutta2012,Gardner1988}. The spiking of Al through 1.5~nm of \ce{SiO2} has also been reported for annealing
        temperature of \SI{300}{\celsius}. \cite{Bierhals1998} Furthermore, void formation by stress release of Al at
        temperatures of 300 to \SI{500}{\celsius} has been reported.  \cite{Hinode1989}

    \subsection{Interfaces}
    To fully characterize the heterostructure, the \textit{interfaces} between the constituent materials have to be
    considered. The physics of interfaces is a broad field with many interesting topics. There are many examples where
    the discontinuity at the interface between two materials can lead to interesting physics, such as Rashba-type
    spin-orbit coupling\cite{Bychkov1984}, or a finite conductivity at the interface between the two insulators
    \ce{LaAlO}/\ce{SrTiO3}\cite{Ohtomo2004}.
    
    The three heterostructures studied in this work have many interfaces, starting from the Si/\ce{SiO2} to the
    metal/oxide interface. Naturally, the addition of an extra layer of material also introduces an additional interface
    where effects might occur.

        \subsubsection{The \ce{Si}/\ce{SiO2} interface}
        The first interface we will consider is the \ce{Si}/\ce{SiO2} interface, where a transition from crystalline
        Si to an amorphous matrix of \ce{SiO2} breaks continuity and translational symmetry. At an atomic scale,
        the crystalline Si and amorphous \ce{SiO2} are incommensurable, save for some theorized phases of \ce{SiO2} that
        would give an abrupt interface.\cite{Pantelides2001}
        
        The Si/\ce{SiO2} interface has been extensively studied in the context of MOSFET technology. The focus here has
        primarily been on defects.  One of the most studied defects at this interface is the paramagnetic
        $\textrm{P}_\textrm{b}$ center\cite{Nishi1972,Thoan2011,Mishima2000} which is linked to surface charges,
        decreased mobility, and the negative-bias temperature instability (NBTI)
        effect\cite{Schroder2003,Campbell2007}. For quantum computation, $\textrm{P}_\textrm{b}$ centers have been used
        in coherent manipulation of spin-dependent charge-carrier recombination of phosphorus
        donors\cite{Stegner2006}. Additionally, the coherence times of these phosphorous donors have been shown to be
        adversely affected in proximity to \ce{P_b} centers.\cite{Paik2010} Defects at the interface have also been
        associated with random telegraph signals at low temperatures and characterized by means of single-electron spin
        resonance \cite{Xiao2003, Xiao2004} and magnetic-field dependent measurements\cite{Prati2006}.

\begin{figure}[htp]
                  \centering
                    \sidesubfloat[]{
                      \includegraphics[width=.45\linewidth]{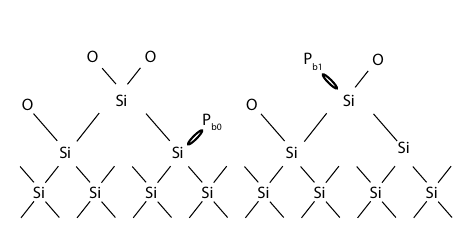}
                      \label{fig:hetero_pbcenter_structure}
                    }
                    \sidesubfloat[]{
                      \includegraphics[width=.45\linewidth]{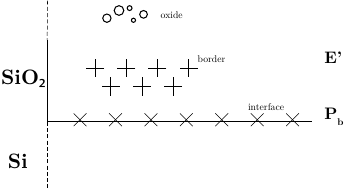}
                      \label{fig:sio2_defects}
                    }
                  \caption{Defects at the Si(100)/\ce{SiO2} interface. \protect\subref{fig:hetero_pbcenter_structure} shows
                    two types of defect at the Si/\ce{SiO2} interface, the \ce{P_{b0}} and \ce{P_{b1}} type dangling bond. \protect\subref{fig:sio2_defects}
                    shows a categorization of defects based on their location relative to the interface. Interface
                    defects are located exactly at the interface between Si and \ce{SiO2}, border defects (such as the
                    E$'$ center) some distance away, and oxide defects are located far away from the interface.}
                  \label{fig:pbcenters_si_sio2}
                \end{figure}

                \paragraph{\ce{P_b} centers, E$'$ centers}
                It is useful to categorize traps of influence at the Si/\ce{SiO2} oxide interface by distance to the
                interface. We can discern, in order of increasing distance, interface traps, border traps, and oxide
                traps (Figure~\ref{fig:sio2_defects}).  
                The two most prevalent defects are \ce{P_b} centers, which is an interface trap, and E$'$ centers, which
                are typically classified as border traps or oxide traps.

                The $\textrm{P}_\textrm{b}$ center is in essence a Si atom with a dangling bond that has formed due to
                the incommensurability of the crystalline Si and the amorphous $\textrm{Si}\textrm{O}_2$. Two types of
                \ce{P_b} center can be further distinguished for Si (100). The $\textrm{P}_\textrm{b0}$ center is
                backbonded to two Si atoms and one oxygen atom, while the $\textrm{P}_\textrm{b1}$ center is backbonded
                to three Si atoms.  The corresponding atomic configurations can be seen in
                Figure~\ref{fig:pbcenters_si_sio2}.  For the $\textrm{P}_\textrm{b0}$ center, the density of interface
                states $D_{\text{it}}$ in the gap has been measured in capacitance-voltage (CV) and electron
                paramagnetic resonance measurements (EPR) to be amphoteric in nature with its maxima at $0.25$ and
                $0.85$ eV above the valence band \cite{Ragnarsson2000}.
          
                The second common defect is the E$'$ center. This defect in \ce{SiO2} is a binary structure: an oxygen
                vacancy with unpaired Si spin on one side, \ce{O3# Si} and a stripped positively charged
                \ce{^+Si# O3} on the other side. The E$'$ center is a common occurrence in thermal oxide and is a possible
                source of oxide fixed-charge.\cite{Caplan1979} It is characterized as a border trap, situated farther
                in the oxide than the \ce{P_b} center.

                The presence of these defects plays a significant part in the formation of quantum dots, and can prevent
                quantum dot formation distorting the confinement potential and providing additional levels for
                tunneling. \cite{Spruijtenburg2013}
                Elimination of defects, such as the $\textrm{P}_\textrm{b}$ center, is commonly achieved through
                chemical passivation by introducing hydrogen in an annealing process. This process reduces the dangling
                bond spatial density from order \SI{e13}{\per\cm\squared} to \SI{e10}{\per\cm\squared} and improves
                numerous parameters related to MOSFET-like transistor operation. It has also been shown to improve
                parameters related to quantum dot formation \cite{Cartier1993,Angus2007}.
                
                In order to create quantum dots reliably, it is thus necessary to perform this annealing step. For
                the heterostructures H1-H3 used in this work, the following applies:
                \begin{itemize}
                 \item For H1, there is no passivation/annealing step. Annealing using a standard forming gas
                anneal of \ce{H2}/\ce{N2} in this architecture was not possible due to dewetting, which is described in
                detail in \cref{sec:dewetting}. The lack of annealing leads to single-hole tunneling through charge
                defects located underneath barriers in all cases \cite{Spruijtenburg2013}.
                 \item  Heterostructure H2 introduces ALD-grown \ce{Al2O3} layers. The covering \ce{Al2O3} layer allows for annealing 
                without dewetting, because it caps the entire structure and contains hydrogen as described in \cref{sec:layer_synthesis}. This
                process is shown to reduce charge defects to a substantial degree \cite{Spruijtenburg2016,Angus2007}.
                 \item Heterostructure H3 should show no appreciable difference at the Si/\ce{SiO2} interface from H2, since
                only the gate-electrode material has been altered. There is the outside chance that the presence of Al
                improves annealing characteristics, in a process commonly known as \textit{alnealing}. 
                \end{itemize}

                \paragraph{Atomically flat/rough silicon}
                As discussed in \cref{sec:si_oxidation}, the interface between silicon and \ce{SiO2} is not expected to
                be atomically flat but rather have a finite roughness. This is further exacerbated by the etching of
                silicon by BHF before high-quality oxide growth, which is known to roughen the surface and leave
                Si\{110\} faceted nanoscale hillocks on the Si surface.\cite{Aldinger2012} For all architectures in this
                work, this interface has been prepared in the same manner, and is expected not to be atomically flat.
                                
                The roughness, with its characteristic length scale, causes the electrochemical potential to fluctuate. If the
                length scale of the carrier wave function is comparable to the length scale of the roughness, this could
                lead to effects such as Anderson localization, a strong decrease in current in the
                device\cite{Svizhenko2007}, or unintentional spinflips \cite{Fleetwood}.
                Optimizing how abrupt the Si/\ce{SiO2} interface is, would be a good candidate for improvement of
                quantum dot properties. As a means for spin-manipulation, the use of atomic steps has also been
                suggested.

                Obtaining atomically flat silicon is possible by using using low-energy ion sputtering and a subsequent 
                temperature step at \SI{700}{\celsius}. \cite{Kim2003} Another method shown to work is to anneal the Si in 
                ultrapure argon at \SI{900}{\celsius} \cite{Kuroda2009}. The surface is heated by Ar bombardment and 
                recrystallizes in the lowest energy state. The key parameter in this process is the amount of trace oxygen,
                which in other cases etches the silicon and induces roughness. Among the reported improvements of this 
                atomically flat Si have been the $1/f$ noise characteristics of MOSFET structures. \cite{Kuroda2009} 
                Chemical means have also been attempted by etching in aqueous \ce{NH4F} (40 \%), although atomics steps were
                not observed.\cite{Aldinger2012}

        \subsubsection{The \ce{SiO2}/\ce{Al2O3} interface - fixed charge in \ce{Al2O3}}
        
        The negative fixed charge $Q_f$ in as-grown \ce{Al2O3} has been measured to be \cite{Dingemans2012a}
        \SI{1e11}{\per\cm\squared}.  Subsequent annealing of this thermally grown layer increased
        $Q_f$ to a maximum of \SI{1e13}{\per\cm\squared}, depending on the annealing temperature and ALD growth
        technique.
        
\begin{figure}[htp]
          \raggedright
            \sidesubfloat[]{
              \hspace{22mm}
              \includegraphics[width=0.5\linewidth]{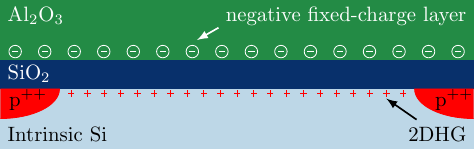}  
              \label{fig:fixedcharge_cartoon}
        }
        
            \sidesubfloat[]{
              \includegraphics[width=.95\linewidth]{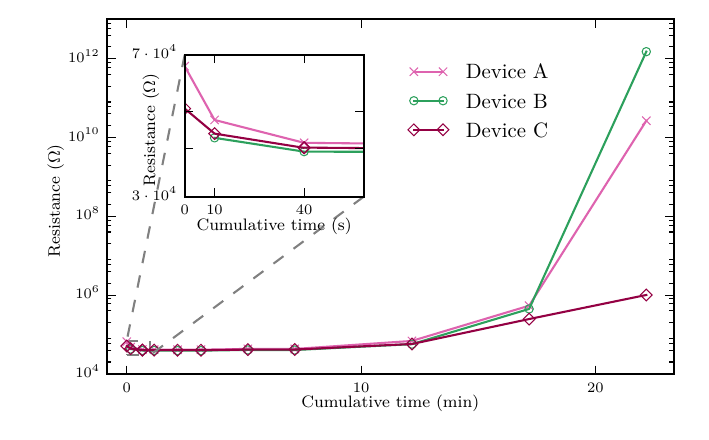}  
              \label{fig:fixedcharge_uvozone}
        }
          \caption{Fixed charge in \ce{Al2O3}. \protect\subref{fig:fixedcharge_cartoon} Schematic drawing of fixed charge, located at the border of the
            \ce{Al2O3}/\ce{SiO2} interface. \protect\subref{fig:fixedcharge_uvozone} Resistance of
            three samples as a function of time exposed in a UV-ozone reactor. Values were obtained by
            measuring the current-voltage relations ($T= \SI{4}{\kelvin}$) and extracting the linear resistance. Inset
            is a detailed view of the behavior for small timescales}
          \label{fig:fixedcharge}
        \end{figure}

        It has been proposed that the negative fixed charge is caused by a very thin interfacial layer that is
        off-stoichiometric\cite{Simon2015} and can be controlled by growing a preceding layer of \ce{HfO2} before the
        growth of the \ce{Al2O3}. Figure~\ref{fig:fixedcharge_cartoon} shows schematically the location of the proposed fixed
        charge $Q_f$.

        Samples without gate structures were grown with \SI{5}{\nm} \SI{250}{\celsius} \ce{Al2O3} layer and another
        layer of \SI{5}{\nm} grown at \SI{100}{\celsius}, and annealed at \SI{400}{\celsius} in \ce{H2}. These samples
        showed hole conduction on the order of $R_{SD}=50$~$\mathrm{k\Omega}$ at $T = \SI{4}{\kelvin}$. Would there be
        no fixed charge present, at \SI{4}{\kelvin} all the charge carriers in the intrinsic silicon are frozen out and
        no conduction would be observed.
        
        An interesting effect on the fixed charge was observed when exposing the samples to ozone in a UV-ozone reactor.
        Figure~\ref{fig:fixedcharge_uvozone} shows the behavior of the resistance of three devices with no gate
        electrodes deposited and only had the layers of \ce{Al2O3} grown. This resistance is shown as a function of
        cumulative time processed in an UV-ozone reactor.  Startup transient effects such as the heating up of the
        mercury lamp in this reactor have not been taken into account. An initial increase of conduction was observed when processing
        the samples in a UV-ozone reactor. This is proposed to be due to further population of the charge traps by
        carriers excited over the band gap by the UV photons. After this initial decrease in resistance it is seen that
        after approximately 10 minutes the resistance increases again. The latter can be explained by the oxidative
        effect of the ozone, which fills the oxide vacancies in the \ce{Al2O3}.
        
        Recently we have used the fixed charge in the \ce{Al2O3} to realize depletion-mode quantum dots in intrinsic silicon\cite{amitonov2017}.

        \subsubsection{The \ce{SiO2}/Metal or \ce{Al2O3}/Metal interface}
        To create the oxide/metal interface, metal is deposited by thermal or e-beam evaporation after lithography. In
        this process, the hot metal vapor is deposited on the colder substrate and interacts with said substrate. Ideally,
        the metal bonds to the substrate and adheres well enough to survive further process steps, such as lift-off. 
        Should adhesion be a problem for the materials involved, an intermediate adhesion layer of a different material
        can be grown. These sticking layers are generally made of small radius atoms, such as Ti.

        Our MOS heterostructures are always dealing with metal/oxide interfaces. Before we continue let us briefly
        consider the stability of phases in compounds consisting of more than two constituent atoms. In our case we are
        dealing with the oxide and the metal. For the \ce{SiO2}/Al interface e.g. we are dealing with the three atomic
        species Si, Al, and O.  Given these three atomic species, the most stable compound is dictated by their Gibbs 
        free energies. The energies of the phases of various materials can be read out from an Ellingham diagram, which 
        tabulates the Gibbs free energy versus temperature.
        
        Given the mixture of Si, Al, and O for H1 at the metal/oxide interface, and the Gibbs free energies of the
        compounds \ce{SiO2} ($\Delta G$ =\SI{-820}{\kilo\joule\per\mol} ) and \ce{Al2O3} ($\Delta G$ =
        \SI{-1015}{\kilo\joule\per\mol}) at $T = \SI{200}{\celsius}$, we expect that the most stable compound is the
        \ce{Al2O3}. This implies that at the interface the Al is reduced by O, which is consumed from the \ce{SiO2}
        layer, leaving elemental Si and \ce{Al2O3}.

\begin{figure*}[]
                          \centering
                            \subfloat{
                          \includegraphics[width=.3\linewidth]{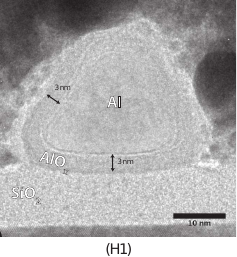}
                          \label{fig:TEM_Al}
                          }
                          \subfloat{
                          \includegraphics[width=.3\linewidth]{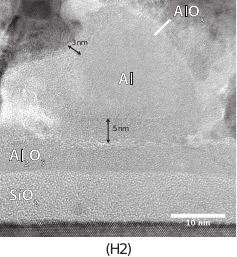}
                          \label{fig:TEM_AlOx}
                        }
                          \subfloat{
                          \includegraphics[width=.3\linewidth]{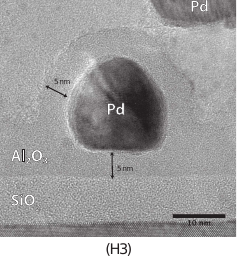}
                          \label{fig:TEM_AlOxPd}
                        }
                        \caption{HR-TEM images of all three heterostructures. (H1) Interaction of metal and oxide at the
                          interface creates an oxide of undefined stoichiometry between the Al and \ce{SiO2}. A slight
                          curvature is observed where the Al ``eats'' into the \ce{SiO2}.  (H2) ALD-grown \ce{Al2O3},
                          with the Al electrode evaporated to be 36~nm thick, followed by Al/Pd (10~nm/60~nm
                          thick). Interaction of Al with \ce{Al2O3} leads to an increased oxide layer thickness of 5~nm, directly
                          under the electrode.  (H3) Pd electrodes show no oxide layer. No contrast difference
                          visible between \ce{Al2O3} grown with ALD at \SI{100}{\celsius} and \SI{250}{\celsius}.}
                          \label{fig:oxideformation}
                        \end{figure*}

        To explore the interfaces, we will now evaluate all three heterostructures on the basis of transmission electron
        microscopy (TEM) studies in Fig. ~\ref{fig:TEM_AlOx}.

        First, let us discuss H1, with its \ce{SiO2}/Al interface, shown in Figure~\ref{fig:TEM_Al}.  An oxide layer
        can be seen surrounding the crystalline Al core. Given that the electrode has undergone the standard oxidation
        in ambient at \SI{150}{\celsius} for 10 minutes, the oxide layer is indeed expected.  However, would this step have
        been the only oxidation process, one would expect more oxide formation at parts of the electrode 
        exposed to ambient. The width of this interfacial layer is also too broad to have originated purely from oxidation
        from the side.  The thickness of the interfacial \ce{Al2O3} layer is determined to be $d \approx \SI{3}{\nm}$.
             
        The interface layer extends beyond the imaginary line of the \ce{SiO2} plane. This indicates that the Al has
        ``eaten'' into the \ce{SiO2} layer.  Previous studies on the phase formed at this interface have concluded that
        a mixture of $\alpha$ and $\gamma$ phases are likely formed. These are less dense and have more ``open''
        structures than the most stable \ce{\alpha-Al2O3} \cite{Roberts1981}.

        Given that the formation of this interfacial layer could contain elemental Si and an undefined phase of \ce{Al2O3},
        it would be beneficial to eliminate the interfacial layer. 
        Heterostructure 2 introduces an intermediate ALD-grown \ce{Al2O3} layer, changing the constituents of the oxide/metal interface as
        compared to heterostructure 1. It would be expected that the interface layer is reduced because the \ce{Al2O3}
        is stable and there is no reason for the Al to be oxidized.         
        Figure~\ref{fig:TEM_AlOx} shows a TEM image of this heterostructure. As compared to heterostructure 1
        the same surrounding oxide layer is observed. However, the interfacial layer has instead increased to
        $d \approx$ \SI{5}{\nano\meter}. It could be that the \ce{Al2O3} underlayer is able to transport oxygen
        efficiently, or the electron affinity of Al allows for easier dissociation than for \ce{SiO2}, combined with a
        diffusion process.

        The third and final heterostructure supplants Al for Pd, and utilizes the ALD-grown \ce{Al2O3} intermediate
        layer. It is immediately visible in Figure~\ref{fig:TEM_AlOxPd} that there is no more surrounding oxide layer, perhaps
        because Pd knows no native oxide. The contrast of the Pd with the surrounding \ce{Al2O3} indicates
        that there is a negligible transition layer.

    \subsection{Morphology}
    In the end, quantum dot formation will be determined by the shape, size, and layout of the gate
    electrodes. Therefore, as a final category we will discuss the morphology of the heterostructure. 
    We will discuss effects related to the morphology of the heterostructure. The shape of the structures as
    they interact with temperature and electric field, change the potential landscape that influences
    quantum dot formation.  
        
    \subsubsection{Thermally induced non-uniform strain} 
        
    Strain is an important parameter in MOSFETs and has been used to manipulate mobility for better
    performance in e.g. finFETs\cite{Neizvestnyi2009,Horstmann2005}.
    Recently it has come to light that the morphology of the gate structure, together with differing
    thermal expansion coefficients, can lead to local strain in the heterostructure. \cite{Thorbeck2014}

    The argument for this is as follows. Measurement of quantum dot devices is nearly always done at
    cryogenic temperatures. The expansion coefficient of each material is different, and thus each
    material contracts at a different rate. The resulting situation for a simple case of a metal gate
    electrode on a semiconductor material is sketched in Figure 1 of Ref.~\cite{Thorbeck2014}. At room
    temperature the materials can be considered nearly stress free, however when cooling down to
    cryogenic temperatures the metal will contract more than the substrate, causing compressive strain in
    the substrate. The strain modulates the CB/VB on both ends of the metal gate, effecting a change in
    potential at these locations. The created potential confines carriers, and an unwanted quantum dot is
    created.
    Mitigating this effect, should one chose to, can be achieved by choosing metals with an expansion coefficient
    comparable to the substrate. In the case of Si, poly-Si would be ideal. Alternatively, the distance to the substrate
    by adding more layers in between electrode and substrate could be increased. This would allow more relaxation of
    stress.
    
    Let us again evaluate the consequences of this effect in our three heterostructures.  Heterostructure
    1 utilizes Al as a gate electrode material, which has one of the biggest thermal expansion
    coefficients. It is separated from the Si by only 8~nm of \ce{SiO2}. One can thus expect that the
    effect will be biggest in this heterostructure.
  
    Because the gate electrode layer is separated by an additional \ce{Al2O3} layer in
    heterostructure 2, it is expected that strain is relieved more before reaching the Si.
    As discussed, mitigating the effect can also be accomplished by changing the electrode material. Pd
    has an expansion coefficient nearly twice as small as Al. It is expected that among the three heterostructures, H3 has 
    the least local strain, firstly because the gate-electrode/Si separation increases by the \ce{Al2O3} layer and secondly 
    it uses Pd as an electrode material.

   \subsubsection{Unintentional dielectric dots} 
   Another morphology effect, concerns the interaction with the electric field, and is also related to the
   formation of unintentional quantum dots.
   
\begin{figure}[htp]
     \centering
       \subfloat{
         \includegraphics[width=0.5\linewidth]{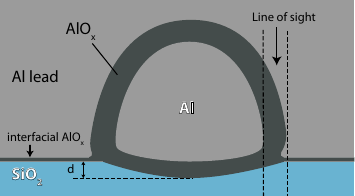}
         \label{fig:sausage_overview}
       }
     \caption{Schematic cross-section of barrier- and lead-electrode. The line of sight for electric field of the
       lead-gate is schematically indicated.}
     \label{fig:sausagedots}
   \end{figure}

\begin{figure}[htp]
\raggedright
       \sidesubfloat[]{
         \includegraphics[width=.45\linewidth]{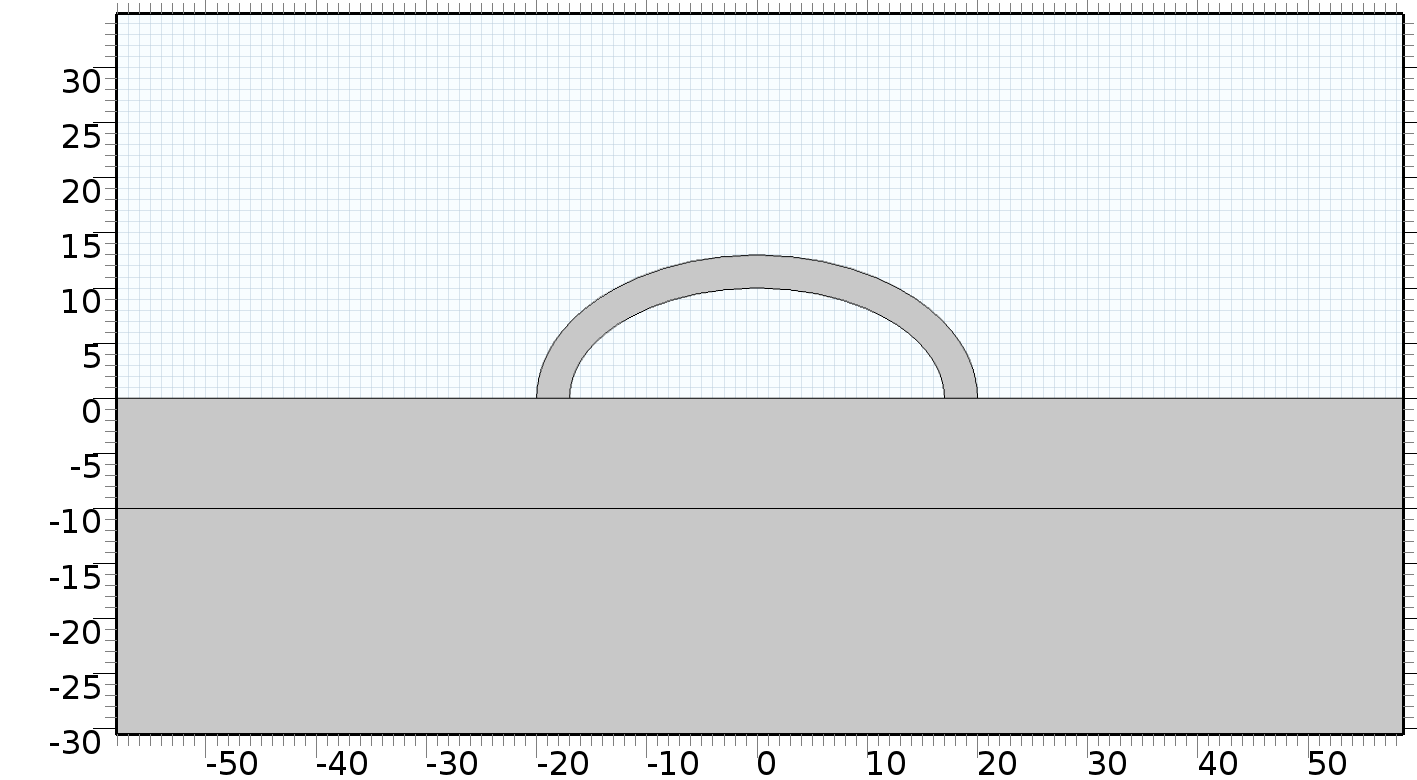}
         \label{fig:sausage_simgeom}
       }
       \sidesubfloat[]{
         \includegraphics[width=.45\linewidth]{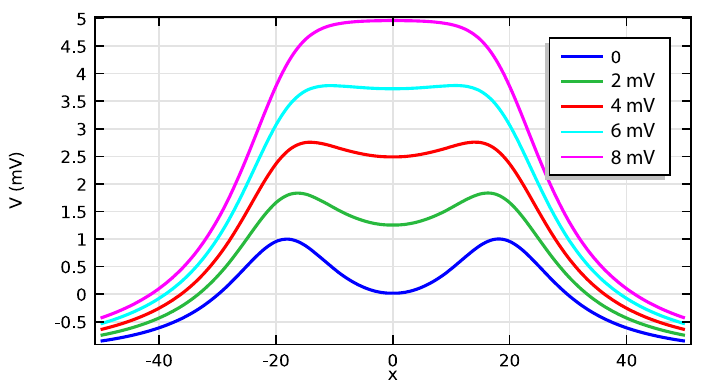}
         \label{fig:sausage_modelsimresult}
       }
     \caption{Simulation of a dielectric dot. \protect\subref{fig:sausage_simgeom} Geometry of
       simulation. \protect\subref{fig:sausage_modelsimresult} Attenuation of the local applied electric field by an increased volume
       of dielectric on either side of the gate electrode as a function of the difference in voltage on
       lead- and barrier-electrode.}
     \label{fig:sausage_sim}
   \end{figure}

   In all heterostructures the first electrode layer is separated by a dielectric from the second electrode layer. This
   has the consequence that, when applying a voltage to the second covering electrode layer, the electric field gradient
   is smaller because it has to traverse more dielectric. This results in a potential that is slightly different at the
   sides of the first layer electrode. Figure~\ref{fig:sausagedots} depicts this schematically with the lead gate having
   a ``line of sight'' to the semiconductor material. Along this line of sight the amount of dielectric material is
   larger.  A simple simulation of this effect is shown in Figure~\ref{fig:sausage_sim}.  Here the resulting potential
   is shown (1~nm below the \ce{SiO2} dielectric) as a function of difference in voltage on
   the lead and barrier gate. The resulting potential deviates from the ideal tunnel barrier potential on the order of
   mV when the voltage on the lead and barrier electrode is equal. This could lead to the formation of an unintentional
   quantum dot. 

   \subsubsection{Grains in aluminum}
   
   \begin{figure}[htp]
     \centering
       \sidesubfloat[]{
       \includegraphics[width=60mm]{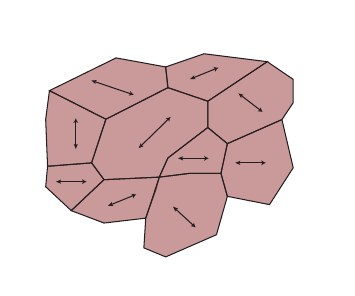}
       \label{fig:grains_schematic}
     }
       \sidesubfloat[]{
       \includegraphics[width=40mm]{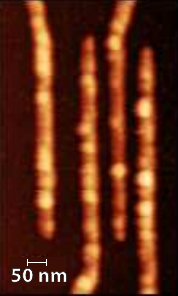}
       \label{fig:grains_afm}
     }
     \caption{Grains in aluminum. \protect\subref{fig:grains_schematic} Depiction of the direction of temperature-induced strain
       in a polycrystalline film. \protect\subref{fig:grains_afm} AFM image of electrodes made from aluminum,
       electron-beam evaporated 36~nm thick. Grain-like features on the order of \SI{50}{\nm} are observed}
     \label{fig:grains}
   \end{figure}

   As discussed in \cref{sec:layer_synthesis}, when evaporating thin films of metal an important parameter is the grain
   size of the resulting polycrystalline film.  Especially when, in evaporating laterally constricted electrodes, the
   grain size becomes comparable to the width of the defined electrodes. This will result in an edge of the film which
   is rough along the line definition. An example of this for Al is given in Figure~\ref{fig:grains_afm}. As in all
   other mechanisms described previously, the resulting potential will thus also be rough. It has been shown that
   roughness can lead to localization of a quantum dot \cite{Voisin2014}.  Another mechanism which could be conceived is
   that the differently oriented strain in separate grains could modulate the electrochemical potential via the
   thermally induced strain mechanism described previously. This is schematically depicted in
   Figure~\ref{fig:grains_schematic}.
 
 \section{The annealing process}\label{sec:annealing}
Elimination and passivation of defects at the Si/\ce{SiO2} interface is commonly achieved by annealing in a \ce{H2}/\ce{N2} environment.

The passivation process for \ce{P_b} centers is governed, in the naive case, by
\begin{align}
  \label{eq:pb_passivation}
  \ch{P_b + H2 & ->[ $k_a$ ] P_bH + H}\\
  \ch{P_bH & ->[ $k_d$ ] P_b + H}
\end{align}

with associated activation energies $E_a = \SI{1.5(4)}{\electronvolt}$ and
$E_d =\SI{2.83(3)}{\electronvolt}$\cite{Stesmans2000}.  The rates of passivation $k_a$ and depassivation and $k_d$ are
determined by temperature, through the Arrhenius expression $k \propto e^{-E_a/RT}$.  The rates of the passivation and
depassivation reaction lead to an equilibrium of the defect density. There is thus an optimum temperature at which the
passivation process is most effective.

In this work we use ALD-grown \ce{Al2O3} to passivate our heterostructure. The process is done in two steps, with the
growth of \ce{Al2O3} taking place at $T_\mathrm{dep}$, and a separate annealing step done at $T_\mathrm{ann}$.  

The existence of hydrogen in ALD-grown oxide was shown through the use of deuterated water as a
precursor.\cite{Dingemans2010} Studying the depth profile of the deuterium, it was determined that the concentration in
the \ce{Al2O3} was reduced, while the content of deuterium in the \ce{SiO2} and at the Si/\ce{SiO2} interface showed a
sharp increase, indicating the deuterium was being consumed in a passivation process.

The degree of passivation is determined by two factors. The first is the oxide growth temperature $T_\mathrm{dep}$, which
determines the density of the oxide, and its hydrogen content. The optimum $T_\mathrm{dep}$ for passivation is reached when the
effusion of hydrogen, controlled by the density of the oxide, is minimized, while the amount of hydrogen present is kept
sufficient to passivate all dangling bonds in the annealing step that follows.

The second factor controlling the degree of passivation is the annealing temperature $T_\mathrm{ann}$. Besides determining the
final equilibrium defect density, $T_\mathrm{ann}$ also determines how quickly this is reached. The best annealing temperature was
previously determined to be \SI{400}{\celsius}\cite{Dingemans2012}.  Using this process, defect densities as low as
\SI{1e11}{\per\cm\squared} have been reported.\cite{Dingemans2015}  The activation energy for passivation has been shown for
ALD-grown \ce{Al2O3} to be $E_a$ = \SI{1.2(5)}{\electronvolt}.

\begin{figure}[htp]
  \centering
    \includegraphics[width=0.7\linewidth]{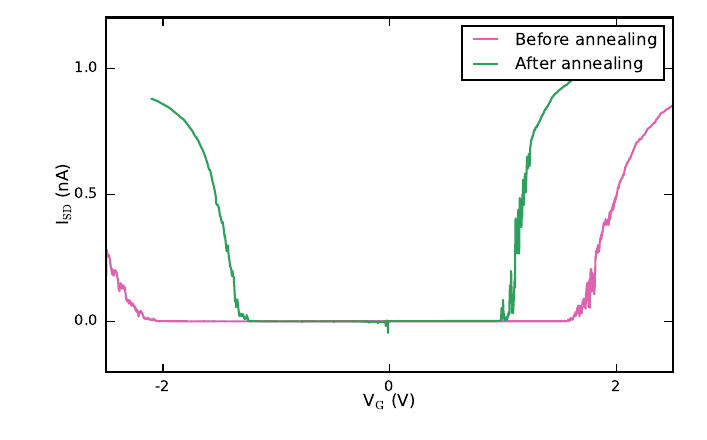}
  \caption{Current-voltage relation of a device as function of voltage on all electrodes $V_\textrm{G}$, both before and after
    annealing a device of the H1 heterostructure in the hole and electron operation regime. $V_{SD} = \SI{1}{\mV}$,
    $T \approx \SI{4}{\kelvin} $}
  \label{fig:passivation_example}
\end{figure}

The ambipolar design of our heterostructures allows us to illustrate this process and its effects quite nicely in the
effects on the threshold voltages $V_{\textrm{Th}}$.  Figure~\ref{fig:passivation_example} shows the current-voltage characteristics of a
device with a heterostructure 2 design, with a single Al accumulation gate. The sample was measured once, before
covering it in ALD-grown \ce{Al2O3} with $T_{\mathrm{dep}}$. It was then annealed at $T_\mathrm{ann} \approx$ \SI{300}{\celsius} in Ar ambient
for 45 minutes, and measured again.  

It can be seen that $V_{\textrm{Th}}$ is reduced for both the hole and electron-operation regime after
annealing. The absolute voltage decrease of $V_{\textrm{Th}}$ between the hole and electron-regime is related to the decrease in \ce{P_b} centers, which
are amphoteric defects capable of storing both negative and positive charge. The layer of \ce{P_b} centers can thus be seen as
an extra capacitor in series with the dielectric. 

\section{Discussion and conclusion}
We have reviewed the many effects that can occur in heterostructures relating to quantum dot formation, and have
identified effects pertaining to three categories: Materials, interfaces and morphology.

The planar quantum dot architecture first introduced by Angus \textsl{et al.}\cite{Angus2007} (H1) has been very
successful. \cite{Lim2009,Veldhorst2015} Further optimization and expansion of this approach therefore bodes very well
for the future. We have identified several areas where improvements can be made.

Firstly, we have observed that the ability to passivate defects at the Si/\ce{SiO2} interface through annealing hinges
primarily on the prevention of dewetting at higher temperatures. In this way, introduction of an ALD-grown \ce{Al2O3}
layer (H2) improves upon the Angus design. This kills two birds with one stone as it prevents dewetting, and enables the
annealing of defects using (at least) the hydrogen present inside the grown film.

Unexpectedly however, an increase of the oxide layer directly beneath the Al electrode is observed. We postulate this
might be due to an enhanced solubility of Al as compared to Si, or an oxygen diffusion process, which would be more
efficient in \ce{Al2O3}.  The ability to anneal these devices now allows very well defined quantum dots up to 180 nm in
length\cite{Spruijtenburg2016}.

Supplanting Al by Pd as the electrode material in H3 prevents the formation of an interfacial layer between the Pd
and the \ce{Al2O3}, as can been seen in TEM images. Pd would reduce temperature-induced strain, since its expansion 
coefficient is less than that of Al. It is unclear how and if the polycrystalline nature of Pd differs from Al, which 
could play a role in strain-related unintentional quantum dots. We have measured many devices with Pd gates, resulting in 
reproducible low-disorder quantum dots \cite{brauns2017}.

A remaining issue is the fixed-charge present in the ALD-grown \ce{Al2O3}. This can be a nuisance for quantum dot
formation, since it induces a 2DHG, even in areas where no electrodes are present.  It can however also be used in
creating depletion dots \cite{amitonov2017}.

Further optimization might include material such as poly-Si or TiN as an electrode material, which are amorphous
materials, and therefore either have a non-uniform or reduced temperature-induced strain. The creation of an abrupt,
atomically flat Si/\ce{SiO2} interface could also improve the reliability of quantum dot formation.

\textbf{Acknowledgements} We acknowledge Alexey Kovalgin and Tom Aarnink for comments on the manuscript and their
technical expertise with regards to atomic layer deposition.  This work is part of the research program “Atomic physics
in the solid state” with project number 14167, which is (partly) financed by the Netherlands Organisation for Scientific
Research (NWO).

\bibliography{siqdcookbook}

\end{document}